\documentclass[cits]{PoS}

\vspace{-4.3cm}
\rightline{\parbox{12cm}{\hfill \rm\small HU-EP-07/41, LU-ITP 2007/02}}
\vspace{2.0cm}

\title{The lattice gluon propagator in stochastic perturbation theory}

\ShortTitle{The lattice gluon propagator in stochastic perturbation theory}

\author{Ernst-Michael Ilgenfritz\\
        Humboldt-Universit\"at zu Berlin, Institut f\"ur Physik, Newtonstr. 15, 12489 Berlin,
        Germany\\
        E-mail: \email{ilgenfri@physik.hu-berlin.de}}
\author{Holger Perlt and \speaker{Arwed Schiller}\\
        Universit\"at Leipzig, Institut f\"ur Theoretische Physik, PF 100 920, 04009 Leipzig,
        Germany \\
        E-mail: \email{Holger.Perlt@itp.uni-leipzig.de}\\
        E-mail: \email{Arwed.Schiller@itp.uni-leipzig.de}
}

\abstract{
   We calculate loop contributions up to four loops to the Landau gauge gluon propagator in
   numerical stochastic perturbation theory. For different lattice volumes we carefully 
   extrapolate the Euler time step to zero for the Langevin dynamics derived from the 
   Wilson action. The one-loop result for the gluon propagator is compared to the infinite volume
   limit of standard lattice perturbation theory.}

\FullConference{The XXV International Symposium on Lattice Field Theory\\
		 July 30 - August 4, 2007\\
		 Regensburg, Germany}

\begin{document}

\section{Introduction}
\vspace{-2mm}

To relate observables measured in lattice QCD to their physical counterpart 
in the continuum, renormalisation is needed.
Besides non-perturbative renormalisation also perturbative approaches are useful.
In addition, it is useful to know as precisely as possible perturbative 
contributions to lattice observables assumed to show confinement properties
in order to separate non-perturbative effects (condensates etc.).
The gluon and the ghost propagator belong to these observables. 

It is well known that lattice perturbation theory (LPT) is much more involved 
compared to its continuum QCD counterpart. The complexity of diagrammatic approaches 
increases rapidly beyond the one-loop approximation. By now only a limited number of results 
up to two-loop accuracy have been obtained.

Applying the standard Langevin dynamics~\cite{Parisi:1980ys,Batrouni:1985jn} 
to the problem of weak coupling expansions for lattice QCD, a  powerful numerical approach for 
higher loop calculations -- called numerical stochastic perturbation theory (NSPT) --
has been proposed in~\cite{Di Renzo:1994sy}.
Amongst other results unquenched $N_f=2$ Wilson loops up to 3-loop order~\cite{Di Renzo:2004ge}, 
plaquettes up to 8-loop order in pure QCD~\cite{Rakow:2005yn}
and renormalisation constants related to the QCD pressure~\cite{DiRenzo:2006nh}
have been calculated.
There is ongoing progress in calculating high-loop perturbative renormalisation 
constants~\cite{Di Renzo:2006wd}.
As a new application we report here on a higher-loop calculation of the perturbative 
contributions to the gluon propagator in Landau gauge. For a similar related study see~\cite{parmaproc}. 
More detailed results will be presented elsewhere.

\section{The Langevin equation}
\label{sec:Langevinequation}
\vspace{-2mm}

The basis of stochastic quantisation is the Langevin equation derived
from the Euclidean action that generates a (quasi-continuous) sequence
of Euclidean field configurations. For gauge theories some special aspects
related to the gauge redundancy have to be taken into account.
Let $t$ the Langevin time, then the Langevin equation reads
\begin{equation}
  \frac{\partial}{\partial t} U_{x,\mu}(t;\eta) = {\rm i}\; \left[
  \nabla_{x,\mu} S_G[U]- \eta_{x,\mu}(t) \right] \;
  U_{x,\mu}(t;\eta) \, ,
  \label{eq:Langevin_I}
\end{equation}
where $\eta = \sum_a \eta^a T^a$ is a random noise field with a
Gaussian distribution
satisfying
\begin{equation}
  \left\langle \eta_{x,\mu}^a(t)\right\rangle_\eta=0\,, \quad
  \left\langle \eta_{x,\mu}^a(t)\eta_{y,\nu}^b(t')   \right\rangle_\eta=
  2 \delta^{ab} \; \delta_{\mu\nu}\; \delta_{xy}\delta(t-t') \, .
  \label{eq:cond_on_eta}
\end{equation}
The notation $\left\langle \ldots \right\rangle_\eta$ denotes an average
over the (external) Gaussian stochastic measure.
As for all Gaussian processes, higher cumulants vanish.
$T^a$ are the (anti-hermitian) generators of the gauge group $SU(N)$.
The differential operator $\nabla_{x,\mu} = \sum_a~T^a\nabla_{x,\mu}^a$ is the left
Lie derivative for any function
on the group and a partial derivative with respect to the links of the lattice.

It can be proven that the gauge fields, in the limit of large $t$ for the
continuous-time Langevin equation, are distributed according to the Gibbs measure
$P[U] \propto {\rm{exp}}(- S_G[U])$.
In practice, the Langevin equation is solved by discretisation of time,
$t= n \epsilon$, with running step number $n$. Therefore, in order to
extract correct physical information, it is not only necessary to go to large
$t$, but also to do the extrapolation $\epsilon \to 0$.
For the solution of the Langevin equation we use the Euler scheme in a way
that guarantees all the links $U_{x,\mu} \in SU(N)$ not to leave the group manifold:
\begin{eqnarray}
  U_{x,\mu}(n+1; \eta) & = & \exp\left({\rm i}~F_{x,\mu}[U,\eta]\right)\; U_{x,\mu}(n; \eta) 
  \label{eq:iteration}
  \\
  F_{x,\mu}[U, \eta] & = & \epsilon~\nabla_{x,\mu} S_G[U] + \sqrt{\epsilon}~\eta_{x,\mu} \, .
  \label{eq:force}
\end{eqnarray}
For $S_G[U]$ we take here the standard plaquette Wilson gauge action.

For the stochastic perturbation theory it is substantial to consider each link
matrix as an expansion in the bare coupling constant $g$. Since $\beta=2N/g^2$,
the expansion reads
\begin{equation}
  U_{x,\mu}(t; \eta) \to 1 + \sum_{l \ge 1} \beta^{-l/2} U_{x,\mu}^{(l)}(t; \eta) \, .
  \label{eq:expansion}
\end{equation}
It simplifies matters if one rescales the time step $\varepsilon = \beta \epsilon$.
Upon the expansion in $g$, the Langevin equation transforms (\ref{eq:iteration})
into a system of simultaneous updates that takes the following form in terms of
the expansion coefficients of $U_{x,\mu}$ (\ref{eq:expansion}) and of the
force $F_{x,\mu}$ (\ref{eq:force})
\begin{eqnarray}
  \label{eq:system}
  U^{(1)}(n+1) & = & U^{(1)}(n) - F^{(1)}(n) \nonumber \\
  U^{(2)}(n+1) & = & U^{(2)}(n) - F^{(2)}(n) + \frac{1}{2} (F^{(1)}(n))^2 - F^{(1)}(n)U^{(1)}(n)  \\
  &\cdots & \, .  \nonumber
\end{eqnarray}
The random noise $\eta$ enters only the lowest order equation through $F^{(1)}$,
the lowest part of the force (\ref{eq:force}) analogous to the expansion
(\ref{eq:expansion}).
Higher orders are stochastic only by the noise fed in from the lower order terms.

A similar expansion like~(\ref{eq:expansion}) exists also for the
(anti-hermitian) vector potential living in the algebra $su(N)$,
\begin{equation}
  A_{x+\hat{\mu}/2,\mu}(t; \eta) \to \sum_{l \ge 1} \beta^{-l/2} A_{x+\hat{\mu}/2,\mu}^{(l)}(t; \eta) \, .
  \label{eq:expansion_of_A}
\end{equation}
Since the vector potential $A_{x+\hat{\mu}/2,\mu}$ is related to the links $U_{x\mu}$ via
$A_{x+\hat{\mu}/2,\mu}  =  \log U_{x,\mu} $,
the separate orders $A^{(i)}$  can be expressed via the orders $U^{(k)}$.
Enforcing unitarity of the originally unexpanded $U_{x,\mu}$ link fields
is tantamount to enforcing anti-hermiticity and tracelessness of all orders
of $A_{x+\hat{\mu}/2,\mu}$.
Whenever we speak about contributions of some order to an observable this has
to be understood in the sense of an expansion
\begin{equation}
  \langle {\cal O} \rangle \to \sum_{l \ge 0} \beta^{-l/2} \langle {\cal O}^{(l)} \rangle \, .
  \label{eq:expansion_of_O}
\end{equation}

\section{The (perturbative) gluon propagator}
\label{sec:gluonprop-formulation}
\vspace{-2mm}

The lattice gluon propagator $D^{ab}_{\mu\nu}(\hat{q})$ is the Fourier transform
of the gluon two-point function, {\it i.e.} the expectation value
\begin{equation}
  D^{ab}_{\mu\nu}(\hat{q}) = \left\langle \widetilde{A}^a_{\mu}(k)
  \widetilde{A}^b_{\nu}(-k) \right\rangle = \delta^{ab} D_{\mu\nu}(\hat{q}) \, ,
\label{eq:D-definition}
\end{equation}
which is required to be color-diagonal. Here $\widetilde{A}^a_{\mu}(k)$
is the Fourier transform of $A^a_{x+\hat{\mu}/2,\mu}$, and $\hat{q}$ denotes
the physical discrete momentum
\begin{equation}
  \hat{q}_{\mu}(k_{\mu}) = \frac{2}{a} \sin\left(\frac{\pi
      k_{\mu}}{L_{\mu}}\right)= \frac{2}{a} \sin\left(\frac{ a q_\mu}{2}\right)
\,, \quad  k_{\mu} \in \left(-L_{\mu}/2, L_{\mu}/2\right]
\label{eq:q-definition}
\end{equation}
corresponding to the integer-valued Fourier momentum 4-vector $k$ on the finite lattice.
Some values $\hat{q}^2(k)$ of the lattice
momentum squared (the lattice equivalent of $q^2$ in the continuum limit)
can be realized by different integer 4-tuples $(k_1,k_2,k_3,k_4)$. 

Assuming reality of the color components of the vector potential
and rotational invariance of the two-point function, the
continuum gluon propagator has the following general tensor structure
\begin{equation}
  D_{\mu\nu}(q) = \left( \delta_{\mu\nu} - \frac{q_{\mu}~q_{\nu}}{q^2}
  \right) D(q^2) + \frac{ q_{\mu}~ q_{\nu}}{q^2}
  \frac{F(q^2)}{q^2} \, ,
  \label{eq:decomposition}
\end{equation}
with $D(q^2)$ and $F(q^2)$ being the transverse and longitudinal propagator,
respectively. The longitudinal propagator $F(q^2)$ vanishes in the Landau gauge.

The lattice gluon propagator $D_{\mu\nu}(\hat{q})$ depends on the lattice
four-momentum $\hat{q}$.
Inspired by the continuum form~(\ref{eq:decomposition}) we consider 
the following lattice scalars
$\sum_{\mu,\nu} \hat{q}_\mu D_{\mu\nu}(\hat{q})\hat{q}_\nu$ and
$\sum_\mu D_{\mu\mu}(\hat{q})$ that should survive the continuum limit.
The first scalar vanishes exactly in lattice Landau gauge.
In this gauge the second scalar function, corresponding
to the tranverse part of the gluon propagator in the continuum limit,
is denoted by
\begin{equation}
  D(\hat{q}) =\frac{1}{3} \sum^{4}_{\mu=1} D_{\mu\mu}(\hat{q}) \,.
  \label{eq:decomposition1}
\end{equation}
On the lattice, this function shows the lower symmetry
of the hypercubic group
in that it depends on the scalar quantities $\sum_\mu \hat{q}^{2n}_{\mu}$, $n=1,2,\dots$
rather than being a smooth function of only $\hat{q}^2$.
Multiplying with $\hat{q}^2$ or $(aq)^2$ we get 
the two versions of dressing functions or form factors
\begin{equation}
  \hat{Z}(\hat{q})=  \hat{q}^2 D(\hat{q}) \,,   \quad    Z(a q)=(a q)^2 D(\hat{q}) \, .
  \label{eq:dressing}
\end{equation}

Using  the expansion (\ref{eq:expansion_of_A})
and $A^{(l)}_{x+\hat{\mu}/{2},\mu}= \sum_a T^a A^{a,(l)}_{x+\hat{\mu}/{2},\mu}$
we obtain the
different {\sl loop} orders $n$ of the perturbative gluon propagator 
(restricted to even powers of $l$ in the sense of (\ref{eq:expansion_of_O})),
\begin{equation}
  \delta^{ab} D_{\mu\nu}^{(n)}(\hat{q}) = \left\langle \,
  \sum_{i=1}^{2n+1}
  \left[ \widetilde{A}^{a,(i)}_{\mu}(k) \,
  \widetilde{A}^{b,(2n+2-i)}_{\nu}(-k) \right] \,
  \right\rangle
  \,.
\end{equation}
Note that already the tree-level contribution,
$D_{\mu\nu}^{(0)}$, arises from the
quantum fluctuations of the gauge fields with $i=1$.
Therefore, the tree-level result for the dressing function, $\hat{Z}^{(0)}(\hat{q})=1$
in the limit $\varepsilon \to 0$ for all sets of lattice momenta
$(k_1,k_2,k_3,k_4)$, is nontrivial and is obtained as the result of averaging.

\section{Practical implementation of NSPT}
\label{sec:implementation}
\vspace{-2mm}

Solving the coupled system of equations~(\ref{eq:system}), one generates
a configuration sequence of expanded gauge fields at finite $\varepsilon$
which can be used to measure the perturbatively constructed observables.
To study the limit $\varepsilon \to 0$, we used $\varepsilon=$ 0.07, 0.05, 0.03, 0.02, 0.01.
It is expected that the autocorrelation time $\tau$ for a chosen observable 
extending over subsequent configurations increases with decreasing $\varepsilon$.
As reasonable compromise between computer time and autocorrelation we have measured the gluon propagator 
after each 20th Langevin step. The remaining autocorrelations are taking into account in the error estimate.

To obtain infinite volume perturbative loop results at vanishing lattice spacing, 
different lattice sizes have to be studied in addition.
We have used $L=6,8,10,12 (16)$ and studied the maximal loop number for the propagator $n_{\rm {max}}=4 \, (1)$.
After reaching equilibrium for the largest Wilson loops to all orders, up to 60000 Langevin steps have been used
to obtain up to 3000 measured gluon propagators.
We checked that expectation values for odd powers of $l$ for all observables indeed vanish in the ensemble average.

The Landau gauge for all considered orders of the perturbative gauge fields ($l_{\rm {max}}=10$) is defined by 
the condition
\begin{equation}
  \sum_\mu \partial_\mu^L A^{(l)}_{x,\mu}=0\,, \quad
  \partial_\mu^L A^{(l)}_{x,\mu}\equiv A^{(l)}_{x+\hat\mu/2,\mu}-A^{(l)}_{x-\hat\mu /2,\mu} 
  \,.
  \label{eq:landaucond}
\end{equation}
For the configurations used in measurements
it is reached by an iterative gauge transformation using the expanded variant of
$ 
U_{x,\mu}^g = G(x) \, U_{x,\mu} G^\dagger(x + \hat \mu) \,.
$
The gauge transformation is chosen as
a perturbative variant of Fourier acceleration~\cite{Davies:1987vs}
with an optimal $\alpha = 1/\hat p^2_{\rm{max}}$
\begin{equation}
  G^{(l)}(x) =\exp \left[\hat F^{-1} \, \alpha \,
  \frac{\hat p^2_{\rm{max}}}{\hat p^2} \,
  \hat F\left( \sum_\mu \partial_\mu^L A^{(l)}_{x,\mu} \right)\right]\,.
\end{equation}
$\hat p^2$ is the non-zero eigenvalues of lattice $- \partial^2$ and $\hat F$ 
($\hat F^{-1}$)
denotes the forward (backward) Fourier transform.
The iterative procedure to reach the Landau gauge stops when, 
for all orders $l$, 
\\
$
(1/V) \sum_x {\rm Tr} \left[
\left(\sum_\mu \partial_\mu^L A^{(l)}_{x,\mu} \right)^\dagger
\left( \sum_\mu \partial_\mu^L A^{(l)}_{x,\mu} \right)
\right] = 0 
$
within double precision.

In the course of the Langevin process 
we mainly follow the prescription given in~\cite{Di Renzo:2004ge}:
after each Langevin step we perform a gauge transformation with
$
G^{(l)}(x)= \exp \left[- \varepsilon \sum_\mu \partial_\mu^L A^{(l)}_{x,\mu}\right]
$
and subtract zero momentum gauge field modes to all orders. 
This keeps the gauge field components finite.

\section{Selected results for the gluon propagator}
\vspace{-2mm}

\begin{figure}[!htb]
  \begin{tabular}{cc}
    \includegraphics[scale=0.60,clip=true]{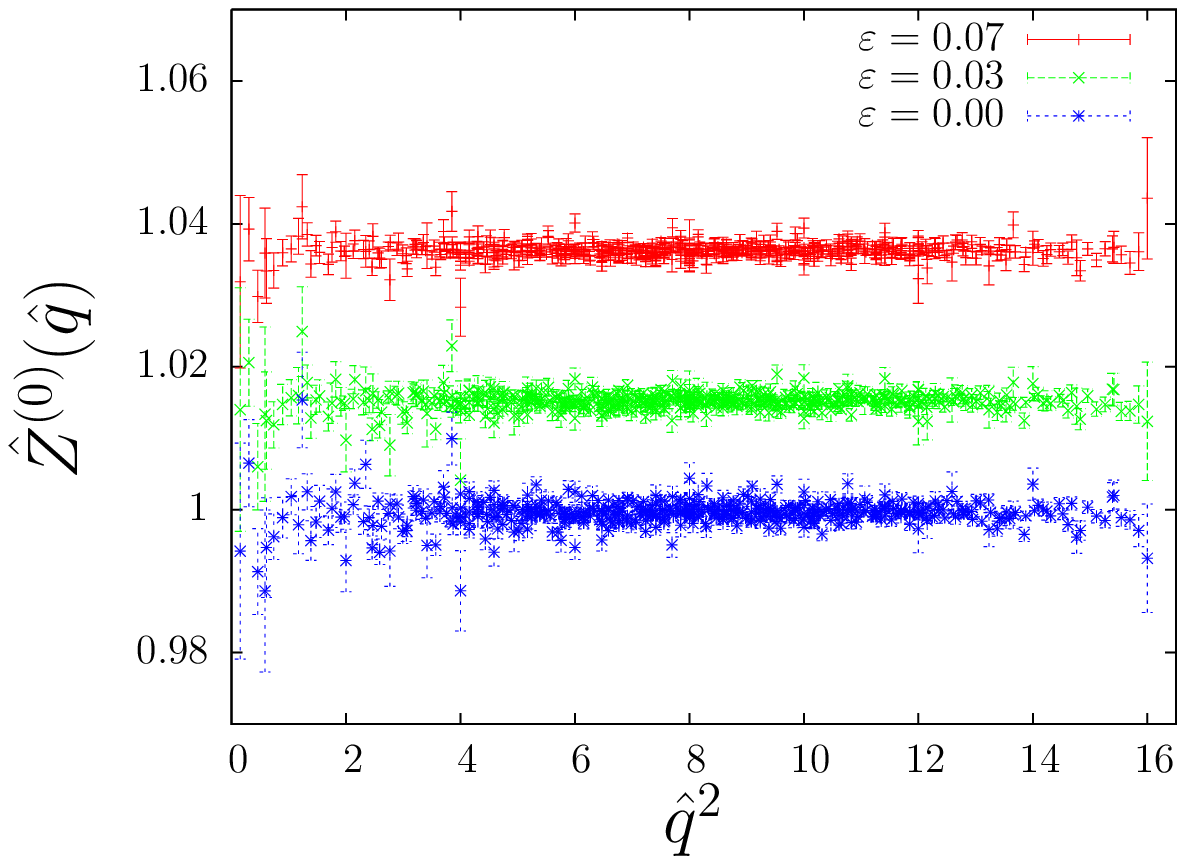}
    &
    \includegraphics[scale=0.60,clip=true]{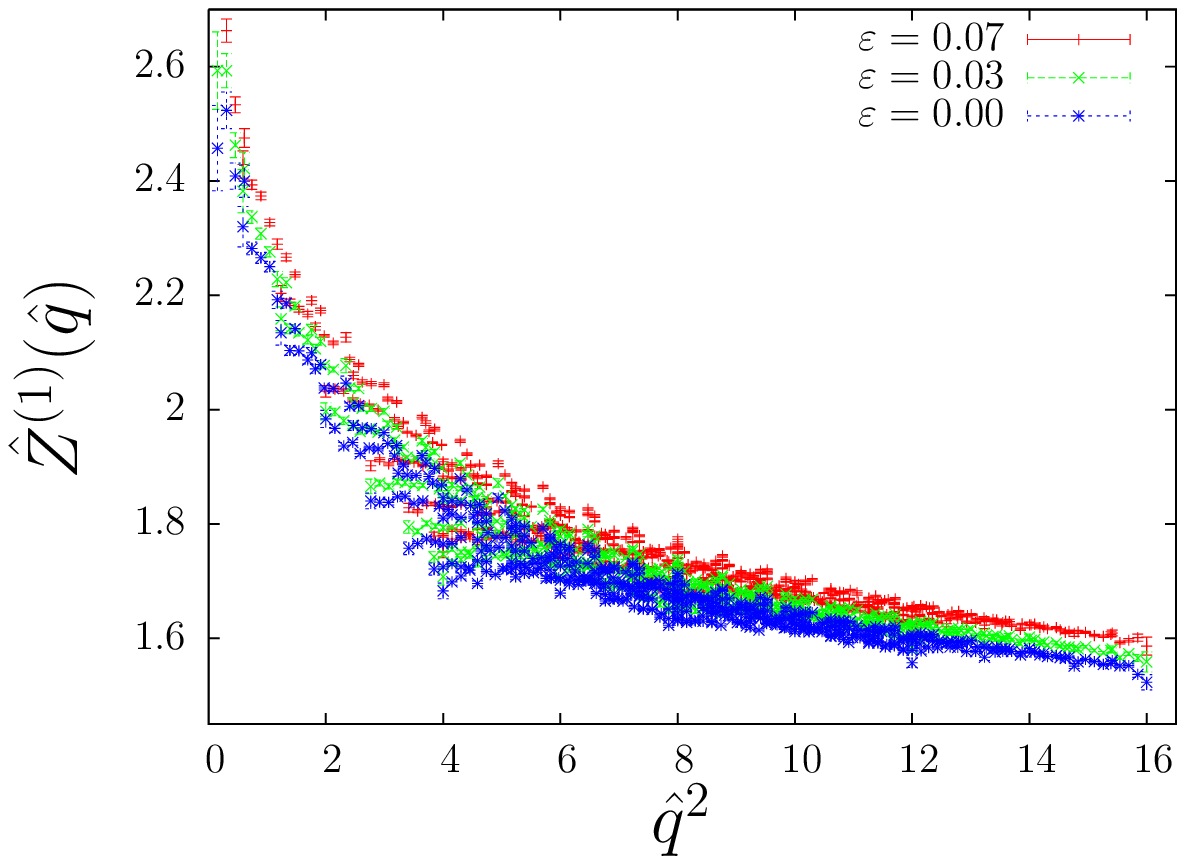}
  \end{tabular}
  \vspace{-5mm}
  \caption{Tree level (left) and one-loop (right) dressing 
  function $\hat Z^{(0,1)}(\hat q)$  vs. $\hat q^2$ at  $L=16$.}
  \label{fig:1}
  \vspace{-2mm}
\end{figure}

The measured gluon propagator data have been averaged over 
equivalent 4-tuples of lattice momenta
and linearly extrapolated to the limit $\varepsilon=0$.
In Figs.~\ref{fig:1} 
we present the extrapolation at lattice size $16^4$ 
for the tree and one-loop dressing function $\hat Z^{(0,1)}$
as function of $\hat q^2$ defined in (\ref{eq:q-definition}),
together with original data
at $\varepsilon=0.07$ and $0.03$ .

\begin{figure}[!htb]
  \begin{tabular}{cc}
    \includegraphics[scale=0.60,clip=true]{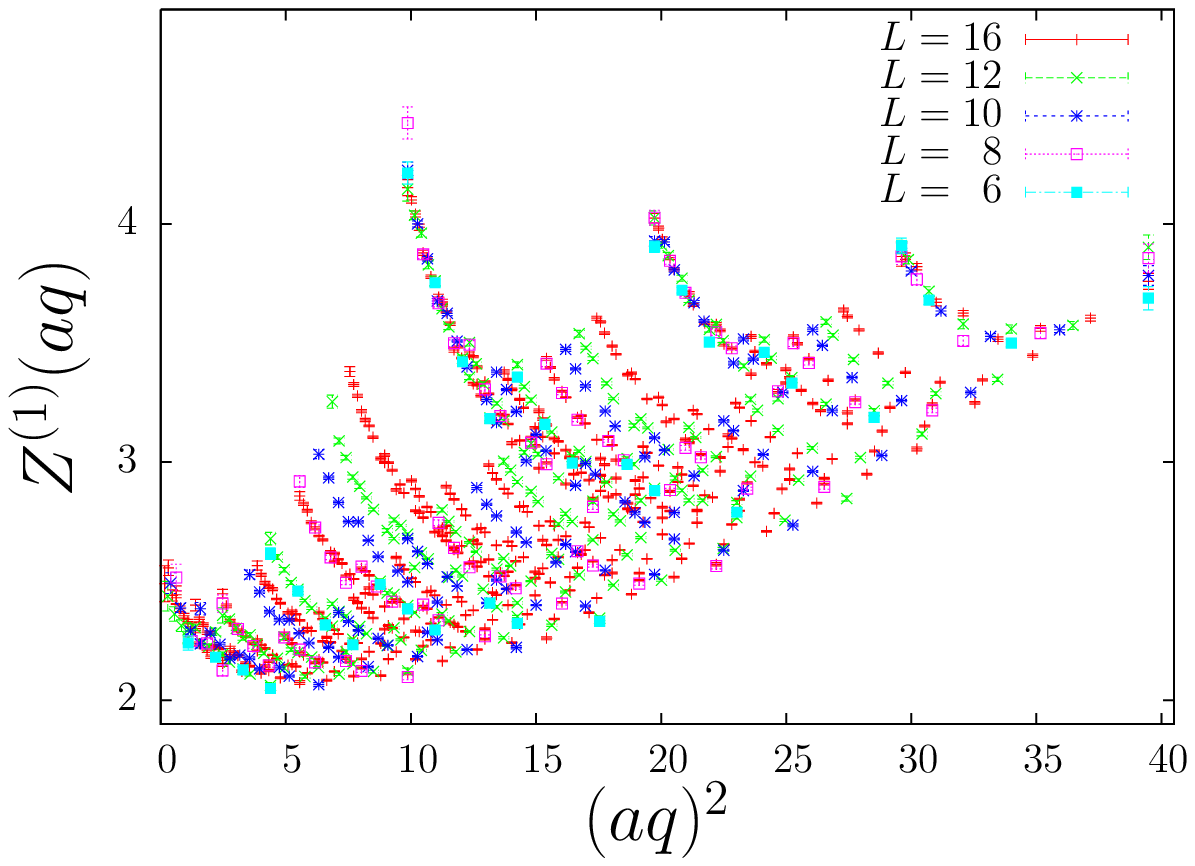}
    &
    \includegraphics[scale=0.60,clip=true]{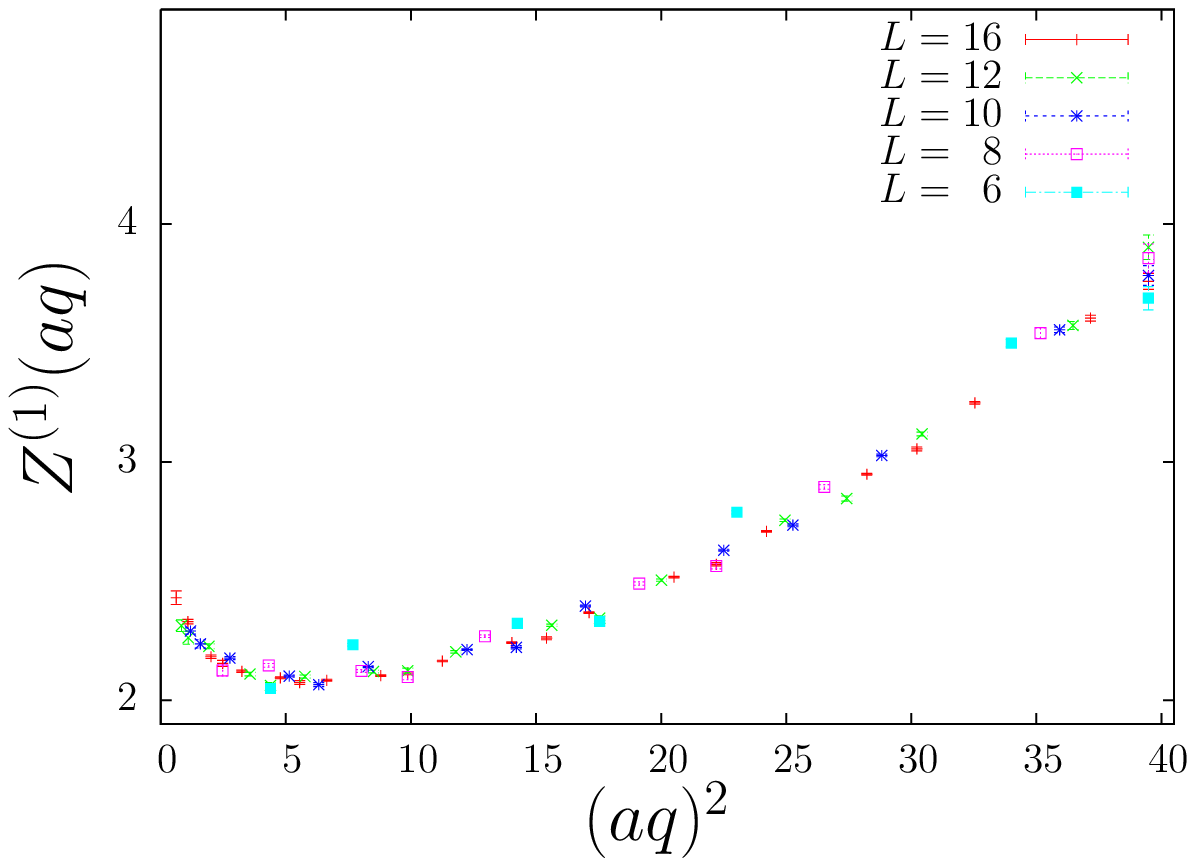}
  \end{tabular}
  \vspace{-5mm}
  \caption{One-loop dressing function $Z^{(1)}(a q)$  vs. $(a q)^2$ at all volumes.
  Left: all independent momentum components. Right: components near the diagonal  
  $(k,k,k,k), (k\pm1,k,k,k), k > 0$.}
  \label{fig:2}
  \vspace{-2mm}
\end{figure}

Figs~\ref{fig:2} 
show a volume and lattice momentum cut dependence using the
other representation of the one-loop dressing function 
$Z^{(1)}(a q)$  vs. $(a q)^2$.
This behaviour is similar in all loop contributions that have been studied.
In the left figure the different branches for off-diagonal momentum tuples are 
clearly seen, which do not possess a continuum limit $a\to 0$.
Restricting to momentum values near the diagonal 
$(k,k,k,k), (k\pm1,k,k,k), k > 0$, a universal
momentum dependence for larger volumes shows up.

\begin{figure}[!htb]
  \begin{tabular}{cc}
    \includegraphics[scale=0.6,clip=true]{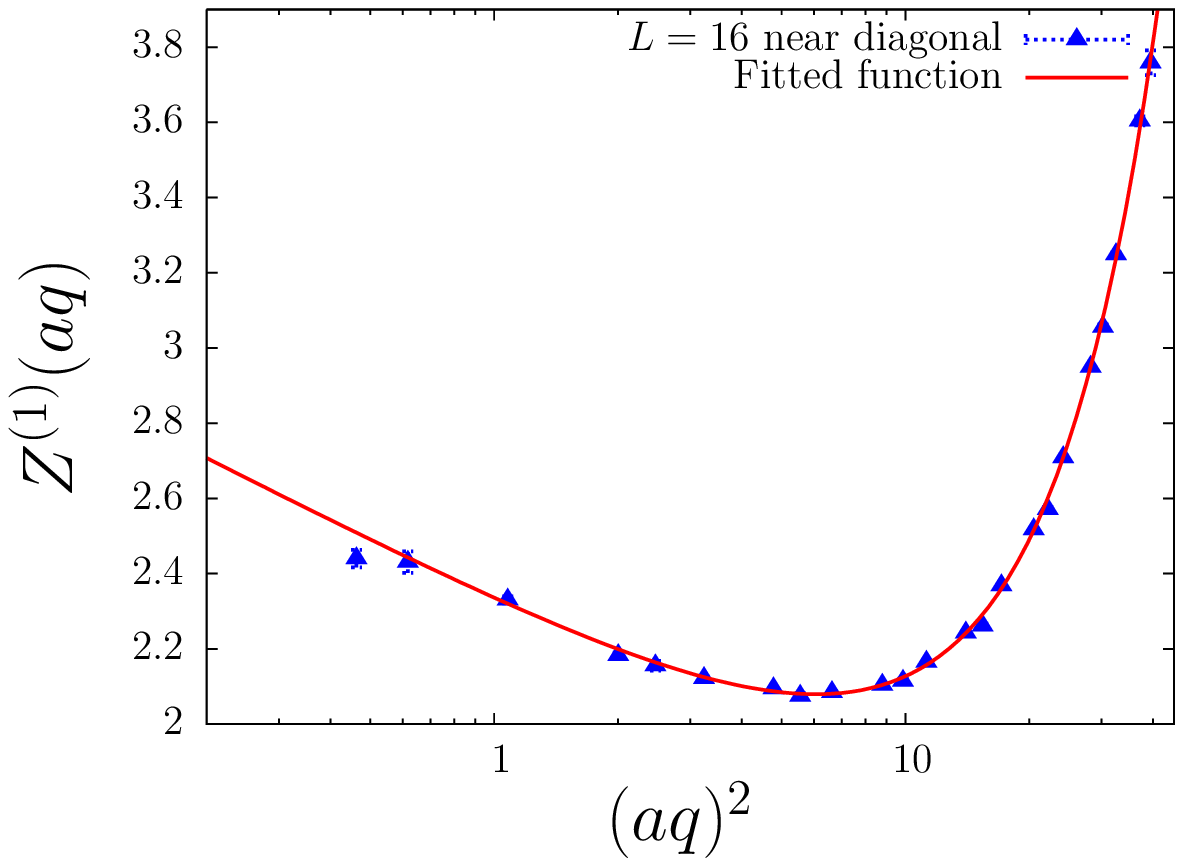}
    &
    \includegraphics[scale=0.6,clip=true]{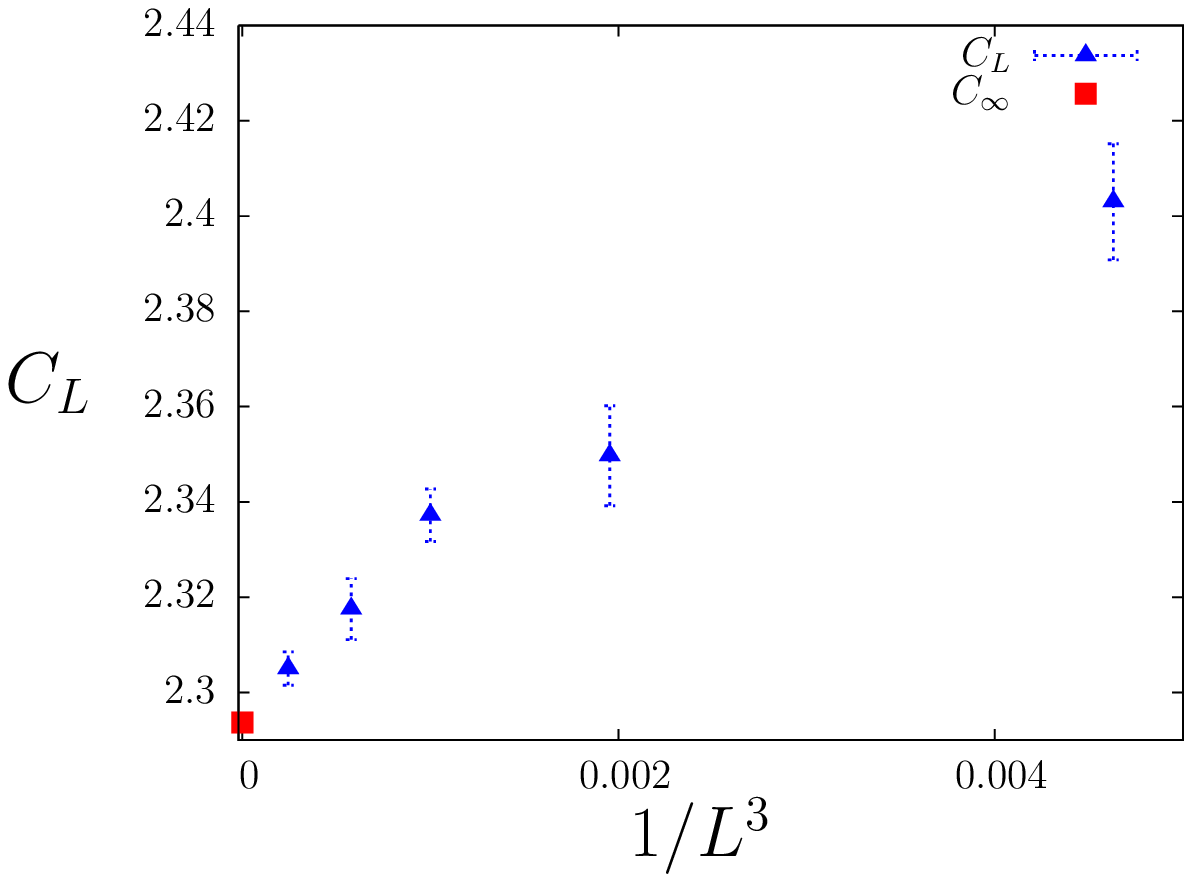}
  \end{tabular}
  \vspace{-5mm}
  \caption{Left: One-loop fitting at L=16, $C_{L=16}=2.3050(35)$.
           Right: $C_L$ vs. $1/L^3$ together with  
	   the known infinite volume result $C_\infty=2.29368$.}
  \label{fig:3}
  \vspace{-2mm}
\end{figure}

This universal curve can be compared with the known one-loop analytic result  
at $L\to \infty$ and $a\to 0$~\cite{Kawai:1980ja}
\begin{equation}
  Z^{(1)} (a q)= - 0.24697 \log (aq)^2 + 2.29368 \, . 
\end{equation}
The aim is to verify the constant 2.29368.
We fit the dressing function {\sl near diagonal} in the form
\begin{equation}
  Z^{(1)}(a q)=-0.24697 \log  (a q)^2  + C_L + c_1  (a q)^2 +c_2 (a q)^4
\end{equation}
by assuming  an universal anomalous dimension 
and parametrise additional lattice artefacts via 
coefficients $c_1$ and $c_2$.
A typical fit at $L=16$ is presented in the left Fig.~\ref{fig:3}. 
The results for different volumes together with the 
known infinite volume result is shown in the figure on the right.
From here  is no doubt that NSPT will reproduce the one-loop analytic result.

\begin{figure}[!htb]
  \begin{minipage}[t]{6.9cm}
     \includegraphics[scale=0.58,clip=true]{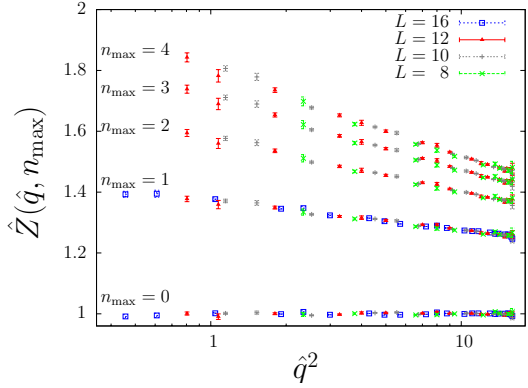}
     \caption{Perturbative dressing function $\hat Z(\hat q,n_{\rm {max}})$ up to four loops 
              (one loop) vs. $\hat q^2$ using $\beta=6$ at $L=8,10,12$ ($16$).}
     \label{fig:4}
  \end{minipage}
  ~ ~
  \begin{minipage}[t]{7.9cm} \vspace{-4.8cm} 
     \section{Summary}

     In the present work we have applied NSPT to calculate the Landau gauge gluon propagator in
     higher-loop perturbation theory. Our results are in good agreement with expectations from
     standard LPT in one-loop. There is a good chance to extract higher loop finite contributions
     using known leading and subleading anomalous dimensions.
     Our results have to be confronted against non-perturbative Monto Carlo results and interpreted.
     The present work is in progress, and we will combine our efforts with those of the Parma group 
     in order to study the perturbative gluon propagator at larger lattices and 
     also the ghost propagator (not discussed here). 
  \end{minipage}
\end{figure}

Finally we present in Fig.~\ref{fig:4} the perturbative dressing function 
$\hat Z(\hat q,n_{\rm {max}})=\sum_{n=0}^{n_{\rm {max}}} \hat Z^{(n)}(\hat q)/\beta^n$  
for near-diagonal lattice 4-tuples summed up to four loops for $L<16$ and up to one loop 
for $L=16$ using $\beta=6$.

\vspace{-0.3cm}
\acknowledgments
\vspace{-0.2cm}
This work is supported by DFG under contract FOR 365 (Forschergruppe Gitter-Hadronen-Ph\"anomenologie).
We acknowledge detailed discussions with Francesco Di Renzo and Christian Torrero.
We thank Paul Rakow for communications, Guiseppe Burgio for help to establish the contact 
with the Parma group and Michael M\"uller-Preussker for his  interest in the progress of this work.

\end{document}